\documentclass[twocolumn,aps,pre,amsmath,showpacs]{revtex4}

\usepackage{graphicx}

\begin{document}

\title{Length scale dependent relaxation in colloidal gels}
\author{Emanuela Del Gado$^{1}$, 
and Walter Kob$^{2}$}
\affiliation{$^{1}$Dipartimento di Scienze Fisiche Universit\`a di Napoli 
``Federico II'', 80125 Napoli, Italy\\
$^{2}$ Laboratoire des Collo\"\i des, Verres et Nanomat\'eriaux, UMR5587 CNRS, Universit\'e
Montpellier 2,
34095 Montpellier, France}
\date{\today}
\begin{abstract}
We use molecular dynamics computer simulations to investigate the
relaxation dynamics of a simple model for a colloidal gel at a low volume
fraction. We find that due to the presence of the open spanning network
this dynamics shows at low temperature a non-trivial dependence on the 
wave-vector which is very different from the one observed in dense 
glass-forming liquids. At high wave vectors the relaxation is due to the 
fast cooperative motion of the branches of the gel network, whereas at low wave 
vectors the overall rearrangements of the heterogeneous structure produce 
the relaxation process.
\end{abstract}
\pacs{82.70.Gg, 82.70.Dd, 64.70.Pf, 67.40.Fd}

\maketitle

In gels, the deep connection existing between their unusual dynamics and the 
open network characterizing their structure is still not understood 
~\cite{exp0,struct}.
In the dramatic slowing down of the dynamics accompanying the gel formation,
the relaxation functions are often stretched and/or, most remarkably, 
compressed, i.e. the time correlators decay faster than an 
exponential~\cite{exp1,exp1_1,estelle,exp2,exp3}, showing a complex dependence 
on length scale. These findings suggest that different relaxation 
mechanisms interplay at a microscopic level, which have not been 
elucidated yet~\cite{cates,puertas}.
In the present paper, we show 
that the formation of the gel network does induce a non-trivial length scale 
dependence of the dynamics in a simple model for colloidal gels. 
We use molecular dynamics computer simulations to study the 
gel formation from the equilibrium sol phase. Our results give evidence that
in the incipient gel, the relaxation at high wave vectors 
is due to the fast cooperative motion of pieces of the gel structure, 
whereas at low wave vectors the overall rearrangements of the heterogeneous 
gel make the system relax via a stretched exponential decay 
of the time correlators. The coexistence of such diverse relaxation mechanisms 
is determined by the formation of the gel network (i.e. the onset of the 
elastic response of the system) and it is characterized by a typical crossover 
length which is of the order of the network mesh size.
This is the first work where such a characterization of
the gel dynamics in colloidal systems has been achieved, thus making important
progress 
%regarding 
as compared to 
previous numerical 
studies~\cite{puertas,zaccarelli,delgado_na,kumar}. 

In colloidal suspensions at low volume fractions gelation 
competes and/or interplays with phase separation. As a consequence, 
coarsening or ordering processes due to the underlying thermodynamics often 
interfere with the gel dynamics. Whereas in the experiments the time scale 
typical of the micro or macro-phase separation is often much longer than the
observation time scales~\cite{exp1,exp1_1}, this is not the case in numerical 
studies using traditional models for colloidal suspensions, where the 
investigation of the gel dynamics has been until now severely 
hindered~\cite{cates,puertas,zaccarelli,delgado_na}.
In order to overcome this problem, 
we have developed a model in which directional 
interactions are able to produce a persistent gel network at relatively 
high temperatures, where phase separation does not occur.  
The model consists of identical particles of diameter $\sigma$, 
our unit of length, that interact via short-ranged two$-$and three body terms. 
The two$-$body term gives rise 
to a narrow well of depth $\epsilon$, our unit of energy, whereas the three 
body term makes that the angle between three neighboring particles is
unlikely to be smaller than 70$^{\circ}$.
As a consequence, at low temperatures there is a competition between 
local structures that are compact (due to the two-body terms) and open 
(due to the three-body terms). 
More details on the potential can be found in 
Refs.~\cite{delgado_kob05a,delgado_kob05c}. 
Differently from other recent models where a fixed connectivity is imposed with the same aim~\cite{zaccarelli},
here it is the balance between the two-body and three-body terms that will
naturally limit the effective functionality of our particles at low 
temperatures, producing an open network with a certain local 
rigidity~\cite{dinsmore}.

\begin{figure}[t]
\includegraphics[width=1.0\linewidth]{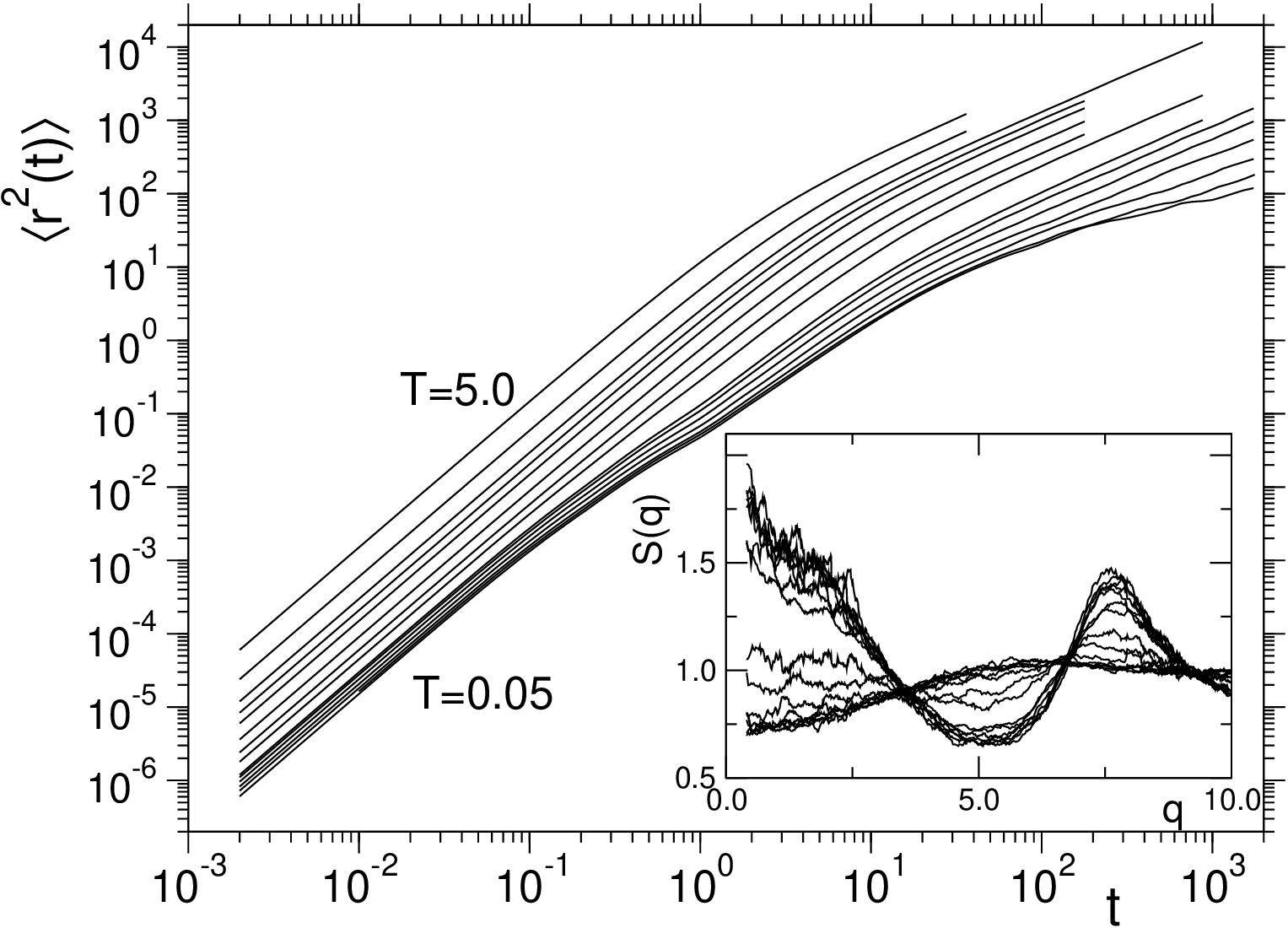}
\caption{
Time dependence of the mean squared displacement of a tagged particle
for all $T$ investigated. Note the presence of a shoulder in
the curves at low $T$ at $t\approx 1.0$ and $t\approx10^2$. Inset:
Static structure factor of the system at $T=5.0,
0.3, 0.2, 0.15, 0.1, 0.09, 0.08, 0.07, 0.05$.}
\label{fig1} 
\end{figure}

With these interactions we have done micro-canonical simulations using
the Verlet algorithm with a time step of 0.002, where time is measured in
units of $\sqrt{m \sigma^2/\epsilon}$, with $m$ as the mass of a particle.
The number of particles is 8000 and the size of the simulation box is
$L=43.09$, which corresponds to a volume fraction of 0.05 (or a particle 
density of 0.1), and the temperatures $T$ are 5.0, 2.0, 1.0, 
0.7, 0.5, 0.3, 0.2, 0.15, 0.1, 0.09, 0.08, 0.07, 0.06, 0.055, and 0.05. 
We carefully equilibrated the system and averaged the results over five 
independent runs. As shown in Ref.~\cite{delgado_kob05a}, there is no sign 
of a phase separation in this temperature range. 
%At this volume fraction
At low $T$ particles are linked by long-living bonds and, via a 
random percolation process, form an open 
network made of chains (particles with coordination number $2$) 
connected by nodes (particles of coordination number $3$). 
The mean value of the chain length distribution between two nodes of the 
network is $10$, giving the typical mesh size.
Such a simple structure of the gel network is a specific feature of this model, 
as compared to other recent studies~\cite{puertas,zaccarelli,delgado_na}, 
making it very useful to investigate the connection between dynamics and 
structural features. 
The static structure factor shows at intermediate and small wave-vectors 
the increase typically found in network forming systems~\cite{exp0} 
(see inset in Fig.~\ref{fig1}). In this $T$-region the relaxation times 
measured from particle diffusion and 
density fluctuations increase with decreasing $T$ quicker than an 
Arrhenius-law. These static and dynamic properties are indeed very similar to
the ones found in real colloidal gels~\cite{exp0,struct,weitz}.

In Fig.~\ref{fig1} we show the time dependence of $\langle r^2(t)\rangle$,
the mean squared displacement (MSD) of the particles for all temperatures 
investigated. 
For high $T$, due to the Newtonian dynamics, one finds at short 
times the ballistic behavior $\langle r^2(t)\rangle \propto t^2$ followed 
at $t\approx 1.0$ by a crossover to
a diffusive behavior, i.e.$\langle r^2(t)\rangle \propto t$. 
At low $T$, apart from the ballistic regime at short $t$, we see at $t\approx
1$ a weak shoulder at around $\langle r^2\rangle \approx 0.05$, i.e. a
localization length around 0.2, due to the onset of the caging regime in 
which a particle is 
trapped by its nearest neighbors. 
At $T=0.05$, more than $97\%$ of the particles belong to the percolating
network. We have separately analyzed the motion of particles 
with different connectivity within the network and found that the motion 
of particles which are weakly connected to the nodes 
(i.e. particles in the middle of a chain connecting two nodes or 
particles belonging to dangling ends) is less affected by this
localization process. 
This indicates that the gel has a 
very flexible structure since the
chains can still perform an 
oscillatory motion without breaking. Thus, the movement of the overall local 
structure makes that the MSD shows only a weak sign of this trapping, 
in contrast to dense glass-forming systems~\cite{kob95}. 
Most remarkably the MSD at low $T$ shows at times $t\approx 10^2$ a 
{\it second} shoulder with a height of around $10^2$, corresponding 
to a localization distance of around 10.
This localization process appears when the spanning network is formed 
and the lifetime not only of the bonds but also of the nodes 
becomes comparable to the longest relaxation time~\cite{delgado_kob05c}. 
It indicates that the system is now viscoelastic over the observation time 
scale and its typical length scale is comparable to the mesh size of the 
disordered network~\cite{footnote1}.
\begin{figure}[t] 
\includegraphics[width=1.0\linewidth]{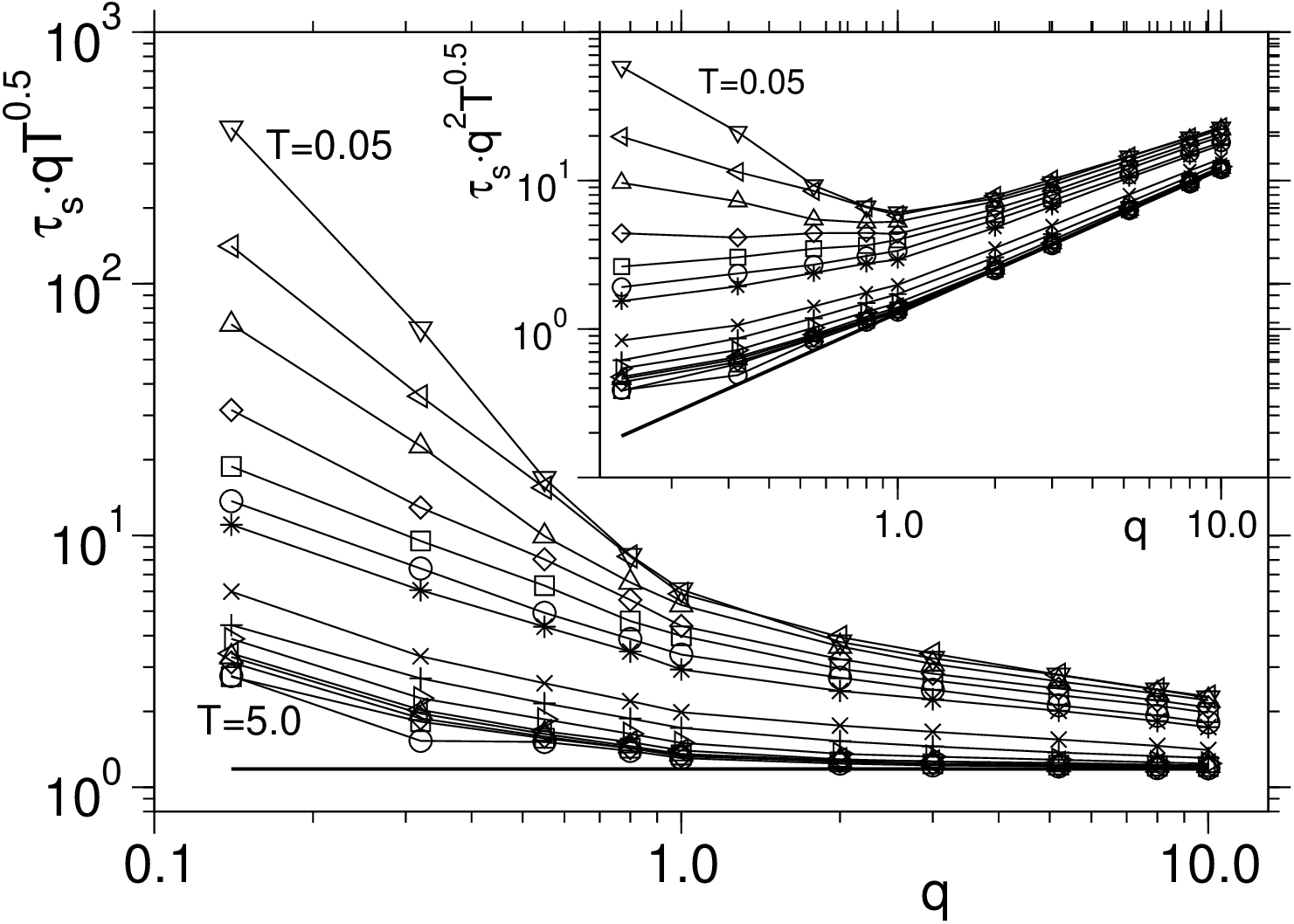} 
\caption{ 
Wave-vector dependence of the relaxation time for all $T$.
Main figure: $\tau_s q \sqrt{T}$ vs. $q$ showing 
the ballistic motion at high $T$ and large $q$. The bold
line is the theoretical expectation for a ballistic motion. Inset:
$\tau_s q^2 \sqrt{T}$ vs. $q$ to check for the onset of the hydrodynamic
regime at small $q$.}
\label{fig2} 
\end{figure}

Relevant information on the relaxation dynamics 
over different length scales
can be obtained from the self intermediate scattering function
$F_s(\vec{q},t)=N^{-1}\sum_{j=1}^N \langle \exp[i \vec{q} \cdot
(\vec{r}_j(t)-\vec{r}_j(0))] \rangle$, where $\vec{q}$ is the wave-vector.
%In Ref.~\cite{delgado_kob05a} we have discussed the $T-$dependence of the
%relaxation time $\tau_s(q,T)$, defined as the time integral of $F_s(q,t)$,
%and shown that {\it for small} $q$ at low temperatures the relaxation times 
%increase quicker than an Arrhenius-law, in agreement with the experimental 
%findings~\cite{weitz}. 
In Fig.~\ref{fig2} we show the $q-$dependence of $\tau_s$ for all temperatures 
investigated. Since at high $T$ the relaxation dynamics at short distances, 
i.e. large $q$, can be expected to be of ballistic nature, it can be 
approximated by the function $F_s(q,t) = \exp[-Tq^{2} t^{2} /(2m)]$ 
and therefore the relaxation time, defined as the time integral of the correlator, should be given by 
$\tau_s= (\sqrt{\pi m /2T})/q$. 
The excellent agreement with our data at high $T$ and large $q$ is shown by 
the plot of the quantity $\tau_s q\sqrt{T}$ in Fig.~\ref{fig2}.
At lower $T$ (but $T > 0.1$) the rescaled data lie higher 
than the ones for high $T$. 
At these $T$ the bond lifetime becomes comparable to the typical 
relaxation time for $q$ values that correspond to distances of the order 
of a few interparticle diameters. Therefore the small aggregates signalled by 
the cluster size distribution~\cite{delgado_kob05a} will have a 
lifetime longer than the relaxation time on this length scale. 
Finally, both at high and intermediate $T$ the data 
deviate at the lowest $q$ where one expects a crossover to the hydrodynamic 
dependence $\tau_s(q,T)\propto q^{-2}$ (see Fig.~\ref{fig2}).
At even lower $T$, the data for large $q$ appear to follow a 
different {\it nearly}-ballistic regime, in that the curves at 
large $q$ are also almost horizontal. We recall that as we lower the 
temperature the bond lifetime is longer than the longest relaxation time in 
the system and that the cluster size distribution
significantly widens~\cite{delgado_kob05a}. 
E.g. at $T\leq0.055$ most of the particles (more than 
$97\%$ at $T=0.05$) belong to one percolating cluster~\cite{delgado_kob05c}. 
This {\it nearly}-ballistic regime at high $q$ thus corresponds to time and length scales 
where the MSD is strongly dominated by the motion of the particles 
connected only weakly to the nodes, i.e. it is due to the 
fast motion of the branches of the gel network.

The length scale $q \simeq 1.0$ marks the crossover to a different 
dynamic regime. In fact, for wave-vectors less than 1.0, i.e. distances 
that are comparable or larger than the length scale of the mesh size, we 
find strong deviations from the previous $q-$dependence 
in that $\tau_s$ increases rapidly with decreasing $q$. 
This strong $q-$dependence, which resembles the one found in dense glass-forming 
liquids, is here observed
at wave-vectors that correspond not to an 
inter-particle distance, but to the mesh size of the network. 
Note that the formation
of the gel network induces a correlation length rapidly increasing
with decreasing $T$, and in fact the hydrodynamic regime at the lowest 
$T$ sets in only at length scales that are larger than the size of the 
simulation box (see inset of Fig.~\ref{fig2}). 
\begin{figure}[t]
\includegraphics[width=1.0\linewidth]{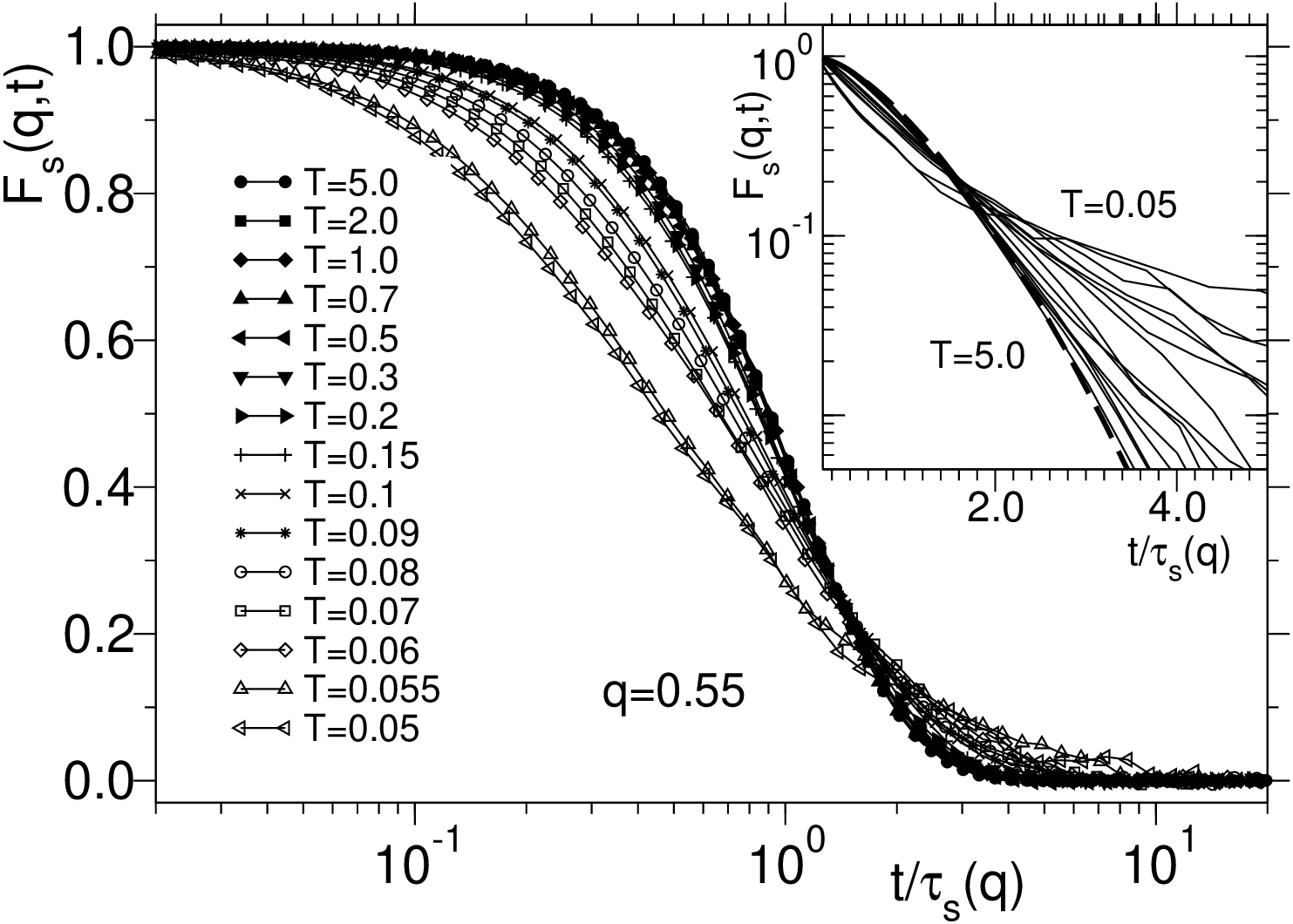}
\caption{
$F_s(q,t)$ vs. $t/\tau_s(q,T)$ for $q=0.55$ and the indicated
temperatures. Inset: The same data in a log-lin representation. The bold
dashed line is a compressed exponential with exponent 1.5.}
\label{fig3}
\end{figure}

We now discuss in more detail the $T-$and $q-$dependence of $F_s(q,t)$, which 
give us further insight in the dynamics.
In Fig.~\ref{fig3} we show $F_s(q,t)$ as a function of $t/\tau_s(q,T)$
for $q=0.55$.
At high temperatures this wave-vector corresponds to the crossover from the
trivial ballistic regime to the hydrodynamic one, i.e. isolated 
particles have a significant probability to collide before making a displacement 
of the order of $2\pi/q$ (leading to an exponential decay of the correlation 
function). The cluster size distribution indicates the presence of small 
aggregates~\cite{delgado_kob05a} but the bond lifetime is smaller 
than the relaxation time on this length scale.
Therefore, the small clusters that move ballistically break up before they have
made a displacement of the order of $2\pi/q$ (which would lead to a Gaussian
decay). Hence the overall relaxation on this length scale is faster
than exponential and basically follows a compressed exponential (CE). 
%with an exponent around 1.5 at $T=5$ (see inset). 
With decreasing $T$ this exponent diminishes monotonously to
around 0.58 at $T=0.05$, giving a stretched exponential (SE) decay. 
This is due to the change of the bond lifetime and of the cluster size 
distribution: If the bond lifetime is comparable to $\tau_s(q,T)$ but the size 
of the aggregates is still small, one expects a nearly exponential
relaxation. With decreasing $T$, the particles aggregate in long-living 
structures and form a percolating
network. Therefore the relaxation curves become more stretched and, at the
lowest temperature considered, can be well approximated by a stretched
exponential. This pronounced stretching can be understood from
the very heterogeneous structure of the network on the length scale of
the size of the mesh.

\begin{figure}[t]
\includegraphics[width=1.0\linewidth]{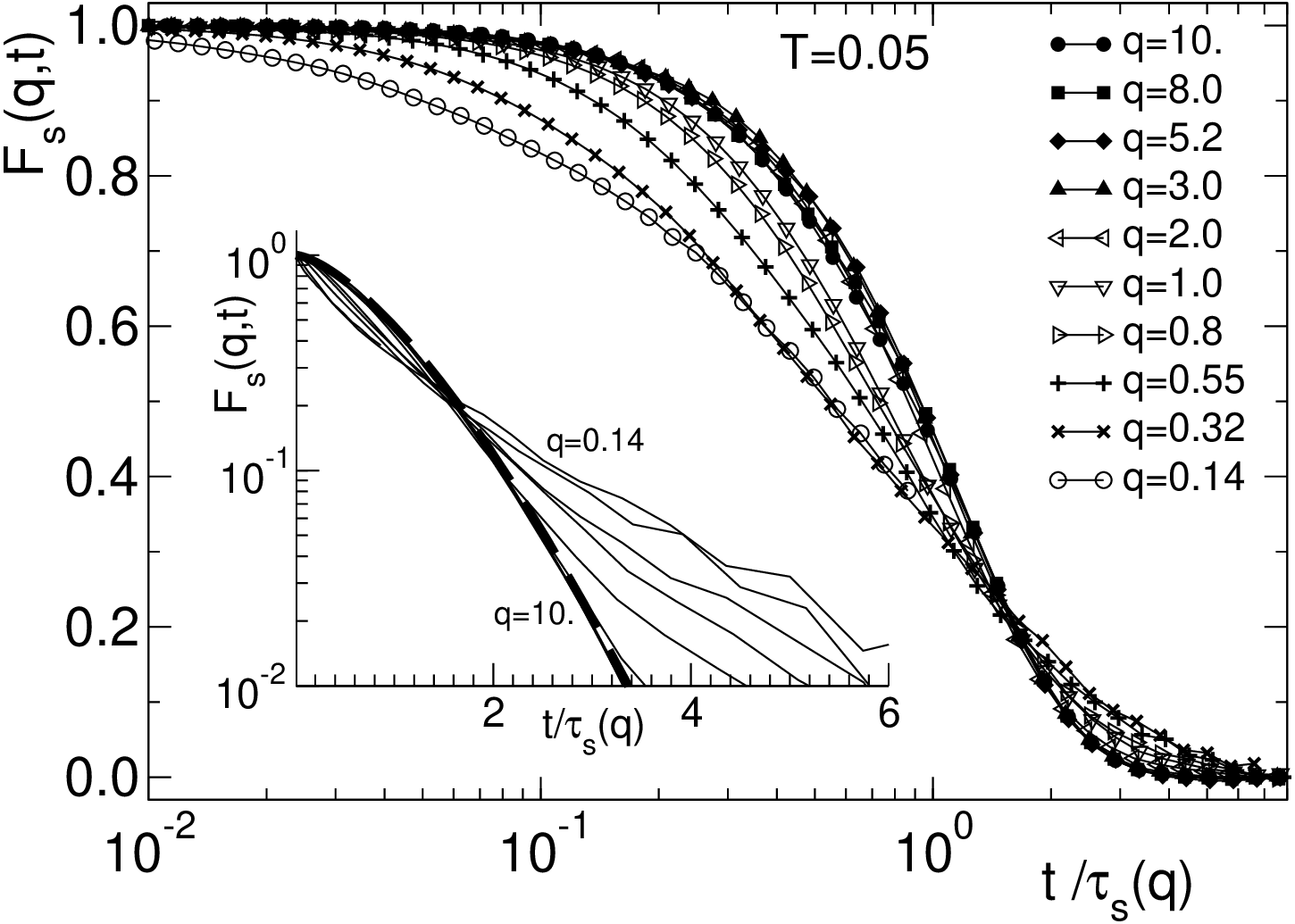}
\caption{
$F_s(q,t)$ vs. $t/\tau_s(q,T)$ for $T=0.05$ and the indicated
wave-vectors. Inset: The same data in a log-lin representation. The bold
dashed line is a compressed exponential with exponent 1.45.}
\label{fig4}
\end{figure}

Let us now analyse $F_s(q,t)$ at the lowest temperature 
$T=0.05$, for different values of $q$, as a function of $t/\tau_s(q,T)$, see
Fig.~\ref{fig4}. 
For large $q$ the curves 
fall all on a master curve and in the inset we show that the latter is well 
described by a CE with an exponent around 1.5. 
At this $T$, finite clusters and free particles 
are extremely rare and the main contribution to the particles MSD comes from 
the particles of the network which are less connected to the nodes. 
This indicates that, for wave vectors $q$ corresponding to a few 
inter-particle distances and smaller, the motion of the branches of 
the percolating cluster leads to a complete decay of $F_s(q,t)$. Hence
their mean free path is larger or comparable than $2\pi/q$ and for these
intermediate and large wave-vectors the motion can be considered
as {\it nearly}-ballistic.
Note that these moving entities do have different masses and therefore 
different thermal velocities. Since $F_s(q,t)$ is the average over the 
different local relaxation functions, assuming that each of them is 
of the Gaussian form given above, it must be expected that $F_s(q,t)$ 
decays more slowly than a Gaussian in time, i.e. that the exponent in the CE is less than 2.0, in 
agreement with our results, see Fig.~\ref{fig4}.

For values of $q\leq 1.0$ the data for $F_s(q,t)$ no longer fall onto a 
master curve and the shape of the correlator becomes a stretched exponential 
(inset of Fig.~\ref{fig4}). Thus only for these intermediate and large length
scales the system shows a relaxation dynamics similar to the one found in
dense glass-forming liquids, i.e. the presence of the disordered structure
on this length scales leads to a heterogeneous dynamics characterized by a
stretched exponential.
We emphasize that the inset of Fig.\ref{fig2} 
well illustrates this peculiar length scale dependence of the relaxation 
dynamics at $T=0.05$, which is crucially determined by the network.
Once that the nodes are persistent enough as compared to the relaxation 
times at small wave vectors (i.e. over distances larger than the network 
mesh size), this network induced relaxation regime sets in: The network 
determines the coexistence of such diverse relaxation mechanisms 
on different length scales. 

Finally, an interesting point which deserves further investigation 
is the relation between the compressed exponential decay of the time 
correlators with the experimental 
observations~\cite{exp1,exp1_1,exp2,exp3}. 
Although this issue is far beyond the aim of the present paper, it is worth 
to recall that here we have approached the gel
formation from the equilibrium sol. 
As a consequence even at the lowest temperature the distribution of
the thermal velocities of the moving entities is the equilibrium one and
hence each contribution to the average $F_{s}(q,t)$ is a Gaussian function
of the time.
In contrast to this, in experiments which study far from equilibrium gels, 
dipolar forces due to 
frozen-in stresses might produce 
a power law distribution of velocities of the moving entities. Hence such 
forces might responsible of the {\it nearly}-ballistic dependence of the 
relaxation time on the wave-vector and of CE decay of time correlators 
with an exponent 
$\sim 1.5$~\cite{exp1,exp1_1,estelle}. Similarly, if the system is sheared and
the perturbation gives (at least locally) rise to a linear displacement
in time one can again expect this type of relaxation~\cite{exp1_1,yamamoto}.
Interestingly, in the non-ergodic gels just mentioned the relaxation is 
supposed to occur via the motion of the connecting branches of the gel 
network and formation or breaking of network nodes, see Ref.\cite{estelle},
as we actually find here. 
These considerations suggest that an analogous analysis of the length 
scale dependence of the dynamics in this model in the far from equilibrium 
gel, obtained by deeply quenching the system, would significantly contribute
to the understanding of the experimental findings.  

In conclusion, our study shows that in colloidal gels at low volume 
fractions the formation of the gel network corresponds to the coexistence 
of very different microscopic relaxation mechanisms. 
Once that the network is formed, fast collective motions of sub-entities 
(branches etc.) of the incipient gel drive the relaxation on small length 
scales, whereas the relaxation on large length scales is due to the overall 
rearrangements of the disordered structure. 
In our model this corresponds to decays of time correlators as diverse as 
CE at high wave-vectors {\it vs.} stretched exponential decays at low $q$. 
The mechanisms discussed here are likely to
affect the stress transmission within the gel network and might therefore
play a role also in the non-ergodic phase investigated in the experiments~\cite{exp1,exp1_1,exp2,exp3}.
We suggest the relaxation mechanisms elucidate here and the length scale 
dependence of dynamics to be a general feature in gels, coupling the 
complex slow dynamics of glassy systems to the structural variety and 
tunability of gelling materials.

Acknowledgments: 
The authors thank L. Cipelletti for many fruitful discussions. 
Part of this work has been supported by the Marie Curie Fellowship
MCFI-2002-00573, the European Community's Human Potential Program
under contract HPRN-CT-2002-00307 DYGLAGEMEM and EU Network Number
MRTN-CT-2003-504712 .


\begin{thebibliography}{99}
\bibitem{footnote1}
This result has implications for the wave-vector 
dependence of the non-ergodicity parameter, i.e. the height of the plateau 
of $F_s(\vec{q},t)$. See also~\cite{puertas,zaccarelli}.
\end{thebibliography}
\end{document}